\documentclass[11pt]{article}

\usepackage[margin=1in]{geometry}
\usepackage{amsmath,amssymb,amsfonts,bm,booktabs,longtable,array,enumitem,graphicx,xcolor,listings,float,tabularx,setspace}
\usepackage[T1]{fontenc}
\usepackage{lmodern}
\usepackage{natbib}
\setcitestyle{authoryear,round,semicolon}
\setlist[itemize]{leftmargin=2em}
\setlist[enumerate]{leftmargin=2em}
\definecolor{codeink}{HTML}{263E4E}
\definecolor{codebg}{HTML}{F5F8F7}
\definecolor{codeframe}{HTML}{CFDCD8}
\definecolor{codekeyword}{HTML}{236B90}
\definecolor{codecomment}{HTML}{52636F}
\definecolor{codestring}{HTML}{526D2D}
\definecolor{codefunction}{HTML}{236B90}
\definecolor{codearg}{HTML}{286B63}

\lstdefinestyle{stlmmR}{
  language=R,
  alsoletter={._},
  otherkeywords={!,!=,~,*,\&,\%/\%,\%*\%,\%\%,<-,<<-,/},
  basicstyle=\ttfamily\small\setstretch{1.0}\color{codeink},
  morecomment=[l]{\#},
  morestring=[b]",
  morekeywords={TRUE,FALSE,NULL,NA},
  keywordstyle=\color{codekeyword}\bfseries,
  commentstyle=\color{codecomment}\itshape,
  stringstyle=\color{codestring},
  identifierstyle=\color{codeink},
  emph={stLMM,iid,gp,nngp,ar1,car,car_time,dagar,dagar_time,car_graph,resid,offset,recover,predict,fitted,get_cor_models,normal,half_t,uniform,as_samples},
  emphstyle=\color{codefunction}\bfseries,
  emph=[2]{formula,data,newdata,graph,car_model,model,variance,n,shrinkage,sub_sample,summary,m,cov_model,ordering,priors,starting,n_samples,chains,chain_control,seed,n_omp_threads,verbose,df,scale,beta,tau_sq,sigma_sq,phi,rho,joint,joint_method,pred_m,pred_ordering,y_samples,return_w_samples,metadata,burn,thin},
  emphstyle=[2]\color{codearg},
  literate={<-}{{\color{codearg}<-}}2,
  backgroundcolor=\color{codebg},
  frame=single,
  rulecolor=\color{codeframe},
  framerule=0.4pt,
  columns=fullflexible,
  keepspaces=true,
  showstringspaces=false,
  breaklines=true,
  xleftmargin=0.75em,
  xrightmargin=0.75em,
  aboveskip=0.8em,
  belowskip=0.8em
}

\newcommand{\by}{\bm{y}}
\newcommand{\bw}{\bm{w}}
\newcommand{\balpha}{\bm{\alpha}}
\newcommand{\bbeta}{\bm{\beta}}
\newcommand{\beps}{\bm{\varepsilon}}

\newcommand{\bu}{\bm{u}}

\newcommand{\bA}{\bm{A}}

\newcommand{\bM}{\bm{M}}

\newcommand{\bQ}{\bm{Q}}

\newcommand{\bW}{\bm{W}}
\newcommand{\bX}{\bm{X}}
\newcommand{\bZ}{\bm{Z}}

\newcommand{\bzero}{\bm{0}}
\newcommand{\trans}{^\top}

\usepackage{letltxmacro}
\LetLtxMacro{\originaleqref}{\eqref}
\renewcommand{\eqref}{Eq.~\originaleqref}

\usepackage{hyperref}
\usepackage{microtype}
\hypersetup{
colorlinks=true, linkcolor=codekeyword, citecolor=codekeyword, urlcolor=codearg,
pdftitle={stLMM: Bayesian Spatial and Space-Time Linear Mixed Models for Small-Area Ecological Estimation},
pdfauthor={Andrew O. Finley} }

\title{\texttt{stLMM}: Bayesian Spatial and Space-Time Linear Mixed Models for Small-Area Ecological Estimation}

\author{Andrew O. Finley\\[0.5em]
\small Department of Forestry\\
\small Department of Statistics and Probability\\
\small Michigan State University, East Lansing, MI, USA\\
\small \href{mailto:finleya@msu.edu}{finleya@msu.edu}}
\date{}

\begin{document}
\maketitle

\section*{Summary}

\begin{enumerate}

\item Ecological monitoring programs often need estimates for areas and periods with few observations. Data may be available as measurements at known locations, measurements identified only by area, or area summaries. These differences in spatial and temporal support require models that connect available observations to the quantities and reporting scales of interest.

\item \texttt{stLMM} is an R package for Bayesian linear mixed models that addresses these needs through a common formula interface for grouped, temporal, point-referenced, areal, and space-time effects, including coefficients that vary across these supports. It accommodates individual measurements and direct estimates with sampling variances, and predicts at unsampled locations or reporting units. Ecological small-area estimation motivates the package, while its model components support broader ecological applications.

\item A shared sparse-precision implementation underlies the model terms. Structured latent effects are collapsed during parameter updates, then recovered for fitted values, diagnostics, prediction, and posterior summaries. This gives users one posterior-draw workflow for propagating uncertainty from model fitting through prediction and aggregation.

\item An illustrative simulation connects point-referenced observations, point observations by area, and direct area estimates to a common prediction target. Timing and memory measurements for larger versions of these settings illustrate how computational demands differ between point-referenced and areal models. The package is also demonstrated with a Washington county biomass example, using national forest inventory data and tree canopy cover to estimate county-year biomass means. An extended article series provides reproducible source code, data, diagnostics, and related model variants.

\end{enumerate}

\noindent\textit{Code and data:} Package source is available at \url{https://github.com/finleya/stLMM}; simulation and timing code and saved timing trials accompany this article as ancillary files.

\emph{Keywords:} Bayesian modeling; ecological monitoring; forest inventory; Gaussian process; R package; small-area estimation; space-time model; spatial statistics.

\section{Introduction}
\label{sec:introduction}

\texttt{stLMM} is an R \citep{RCoreTeam2024} package for fitting Bayesian linear mixed models with structured spatial, temporal, and space-time random effects. The package provides a common formula-based interface for models whose latent terms are indexed by groups, times, locations, areas, or combinations of these supports. Implemented terms include independent and identically distributed (iid) grouped effects, autoregressive (AR) temporal effects, Gaussian process (GP) and nearest-neighbor Gaussian process \citep[NNGP,][]{Datta2016, Finley2019} point-referenced effects, conditional autoregressive (CAR) and directed acyclic graph autoregressive \citep[DAGAR,][]{DattaDAGAR2019} areal effects, separable areal space-time effects, and structured varying coefficients.

These terms are designed to accommodate dependence structures commonly encountered in ecological data, whether responses are continuous measurements, binary observations, or counts. For example, GP/NNGP terms can describe spatial gradients across landscapes, CAR/DAGAR terms can represent similarities among neighboring habitat or management units, and AR terms can capture temporal persistence and the decay of dependence between surveys.

The package was developed to support prediction of space- and time-indexed ecological variables, particularly in monitoring programs where users increasingly require estimates for spatial, temporal, and biophysical domains finer than those targeted by the original sampling designs. Large-scale ecological monitoring programs provide essential information about environmental status and change, but their designs are often optimized for broad reporting domains. As a result, many user-defined domains contain too few observations to support reliable direct, design-based estimates \citep{prisley2021, Dumelle2022, Knott2023}. Model-based small-area estimation (SAE) methods address this problem by combining sparse observations, auxiliary information, and statistical models that borrow strength across related units or domains \citep{RaoMolina2015}.

SAE methods are often described as either area-level or unit-level models. In area-level models, the response is a direct estimate for a reporting domain, and the model accounts for its sampling variance while relating the estimate to domain-level covariates. The Fay-Herriot (FH) model is the canonical area-level model \citep{FayHerriot1979}. In unit-level models, the response is measured on sampled units, such as plots, and domain summaries are obtained by predicting over the target population or support and then aggregating. The Battese-Harter-Fuller model is the canonical unit-level model \citep{BatteseHarterFuller1988}.

These classical models provide useful reference points, but modern ecological applications often require combinations of model components that extend beyond either formulation alone. Examples include point-referenced plot effects, areal effects over reporting domains, temporal and space-time dependence, residual-variance models for direct estimates, structured varying coefficients, prediction to unsampled supports, and posterior summaries over user-defined reporting domains.

\texttt{stLMM} was written for these broader model-building and prediction problems. Fixed effects are specified using standard R formula syntax, while special model terms add iid random effects, structured latent processes, and residual-variance models. The package emphasizes latent terms that can be represented by covariance functions or sparse precision matrices using a unifying design, allowing users to specify dependence induced by distances, neighborhoods, adjacency relationships, or directed acyclic graph orderings. This formulation gives users a single interface for fitting models that borrow strength across sampled units, reporting domains, space, and time.

The package is designed for both direct-estimate and unit-level workflows. A response may be a domain-level direct estimate or an observation measured at a sampled unit; latent processes may be point-referenced, areal, temporal, or space-time; and fitted models may be used to predict at new locations or retained missing-response supports before users aggregate or otherwise transform posterior draws into reporting-domain summaries. Rather than treating these cases as separate software problems, \texttt{stLMM} represents them as combinations of model terms within a shared Bayesian linear mixed-model framework.

\section{Modeling framework}
\label{sec:model-representation}

The package uses a latent mixed-model representation whose computation is organized around sparse precision matrices. For Gaussian responses, the basic observation model is
\begin{equation}\label{gauss_latent_model}
  \by = \bX\bbeta + \bZ\balpha + \bA\bw + \beps,
\end{equation}
where $\bX$ is the fixed-effects design matrix with associated regression coefficients $\bbeta$, $\bZ$ is the random-effects design matrix for explicit iid grouped effects $\balpha$, $\bA$ maps the stacked vector $\bw$ of random effects from structured latent process terms to observations, including covariate scaling for structured varying-coefficient terms, and $\beps$ is the residual component. The iid grouped effects have conditionally independent Gaussian priors and are sampled directly. They are therefore represented separately from the structured latent processes in $\bw$, which are integrated out during fitting and recovered after fitting when needed.

The default residual model is homoskedastic,
\begin{equation}
  \beps \sim N(\bzero, \bW^{-1}),
\end{equation}
where $\bW = \mathrm{diag}(p_1,\ldots,p_n)$ and $p_i=1/\tau^{2}$. Other options allow fixed row-specific precisions, such as $p_i=1/\widehat v_i$ for direct-estimate residual variances, and sampled residual-variance models, such as $p_i=1/\tau_i^2$, with additional options described in the package documentation.

For binomial and negative binomial responses, \texttt{stLMM} uses P{\'o}lya-Gamma data augmentation to recast the likelihood as a conditionally Gaussian problem with a diagonal working precision \citep{Polson2013,WindleDynamic2013,Zhou2012}. Binary probit models use Albert--Chib latent-normal augmentation with fixed unit latent variance \citep{AlbertChib1993}. These augmentations allow the same linear predictor in \eqref{gauss_latent_model}, and the same sparse precision machinery, to support Gaussian, binary, and count responses.

Each structured process term defines a latent support, such as locations, times, areas, or area-time combinations, and contributes a prior precision to the stacked process vector $\bw$. Point-referenced terms describe dependence among locations through their coordinates, whereas areal terms describe dependence among areas through a neighborhood graph. Temporal terms extend the representation to dependence across observation times. Structured varying coefficients allow associations with environmental predictors to vary across these supports. For one process term with latent support $s_1,\ldots,s_{q_k}$, the finite latent vector is
\begin{equation}
  \bw_k = \left(w_k(s_1), \ldots, w_k(s_{q_k})\right)\trans .
\end{equation}
With $K$ structured process terms,
\begin{equation}
  \bw = \left(\bw_1\trans,\ldots,\bw_K\trans\right)\trans,
  \qquad
  \bw \sim N(\bzero, \bQ_w^{-1}),
  \qquad
  \bQ_w = \mathrm{blockdiag}(\bQ_1,\ldots,\bQ_K),
\end{equation}
where $\bQ_k$ is the prior precision for the $k$th structured process term, conditional on its covariance or precision parameters. This block-diagonal prior represents conditional independence among distinct structured process terms, while dependence among observations is induced by mapping the latent processes to the data through $\bA$.

Given the process map $\bA$ and observation precision $\bW$, the key sparse matrix associated with the structured latent process is
\begin{equation}
  \bM = \bQ_w + \bA\trans \bW \bA.
\end{equation}
This matrix shows how prior structure from the process precision, information from the likelihood, and the observation-to-process mapping combine in one sparse linear-algebra problem. Here $\bA$ is the matrix for likelihood-contributing rows and $\bW$ is the corresponding diagonal observation-precision matrix; missing-response rows may remain in the fitted design and latent support, but they do not add likelihood information.

The Markov chain Monte Carlo (MCMC) sampler integrates $\bw$ out during parameter updates, as in the collapsed Gaussian spatial models of \citet{FinleyBanerjeeGelfand2015SpBayes,Finley2019}. Gaussian models recover latent process draws after fitting; augmented non-Gaussian models draw them within the chain for the next augmentation update and retain them for prediction and summaries. The collapsed updates target the fixed effects, iid grouped effects, residual-variance parameters where applicable, and process covariance or precision parameters. This provides a common inferential engine for point-referenced, temporal, areal, and area-time terms. Derivations and implementation details are provided in the package's \href{https://finleya.github.io/stLMM/articles/a12-modeling-software-details.html}{technical reference}.

This shared computational structure is based on sparse precision assembly and sparse Cholesky factorization of $\bM$. Structured process terms enter the model through covariance functions or precision matrices, with most scalable terms represented through sparse precisions induced by neighbor sets, adjacency relationships, temporal ordering, or directed acyclic graph orderings. Once the formula terms and latent supports are defined, the sparsity pattern is fixed, allowing efficient reuse of the symbolic Cholesky factorization throughout sampling. Matrix values and numerical factorizations are updated as parameters change. The sampler uses these shared linear-algebra operations across spatial and space-time models.

Sparse matrix operations rely on CHOLMOD through the R \texttt{Matrix} package's \texttt{C} API \citep{ChenEtAl2008CHOLMOD,MatrixPackage}, while dense operations use the BLAS and LAPACK libraries linked to R \citep{BlackfordEtAl2002BLAS,AndersonEtAl1999LAPACK}. The sampler treats the factorized $\bM$ as an operator: collapsed likelihood evaluations, Gaussian coefficient updates, and latent-process recovery are expressed through CHOLMOD-backed solves, sparse matrix-vector products, and inner products. These collapsed calculations avoid explicitly forming $\bM^{-1}$ or the marginal observation covariance and its inverse.

Building on CHOLMOD, BLAS, and LAPACK is an important part of the package design because it keeps the computational engine flexible and extensible. These libraries continue to benefit from improvements in hardware, CPU and GPU support, and multithreading. CHOLMOD automatically selects fill-reducing orderings, including approximate minimum degree (AMD) and, when warranted and available, METIS \citep{KarypisKumar1998METIS}; users can override this choice through \texttt{cholmod\_control}.

\section{Package interface and workflow}
\label{sec:package-overview}

The user interface is deliberately small. Models are fit with \texttt{stLMM()}, structured latent processes are recovered with \texttt{recover()}, and posterior predictions are generated with \texttt{predict()}. Standard S3 methods are provided for printing, summaries, fitted values, posterior-sample conversion, and diagnostic plots. A typical workflow is as follows.

\noindent
\begin{minipage}{\linewidth}
\begin{lstlisting}[style=stlmmR]
fit <- stLMM(formula, data = dat, ...)
rec <- recover(fit, ...)
pred <- predict(rec, newdata = newdat, ...)
\end{lstlisting}
\end{minipage}
\par

For models with only fixed effects and iid grouped random effects, fitted values and predictions can be obtained directly from the fitted object. For models with structured spatial, temporal, or space-time process terms, \texttt{recover()} prepares latent process draws for fitted values, diagnostics, and prediction. Posterior draws can be subsampled during recovery and prediction, allowing users to control the size of downstream objects.

The main model terms are
\begin{itemize}[noitemsep]
\item \texttt{iid()} for independent and identically distributed grouped random effects;
\item \texttt{gp()} for dense Gaussian process effects on point-referenced supports;
\item \texttt{nngp()} for nearest-neighbor Gaussian process effects on point-referenced or space-time supports;
\item \texttt{ar1()} for autoregressive temporal effects;
\item \texttt{car()} for conditional autoregressive areal effects;
\item \texttt{dagar()} for directed acyclic graph autoregressive areal effects;
\item \texttt{car\_time()} for separable CAR-time effects;
\item \texttt{dagar\_time()} for separable DAGAR-time effects.
\end{itemize}
With the exception of \texttt{gp()}, these process terms are represented through sparse precision matrices.

Formula interactions turn these terms into group-varying or space-time varying effects. For example, varying effects for a generic covariate \texttt{x} can be written as \texttt{x:iid()}, \texttt{x:gp()}, \texttt{x:nngp()}, \texttt{x:car()}, \texttt{x:dagar()}, \texttt{x:car\_time()}, or \texttt{x:dagar\_time()}. This makes structured varying coefficients part of the same formula grammar used for random intercepts.

Point-referenced spatial and space-time data are handled using the \texttt{gp()} and \texttt{nngp()} terms. Covariance functions are discoverable through \texttt{get\_cor\_models()}; current spatial options include the Mat{\'e}rn and exponential, and space-time options include separable and nonseparable models. Any number of coordinate columns can be passed to define the Euclidean space used for constructing the covariance. For space-time covariance models, coordinate order is part of the model, with the final coordinate treated as time. The dense \texttt{gp()} term is intended for smaller supports, whereas \texttt{nngp()} provides the scalable point-referenced option.

Areal spatial and space-time data are handled using \texttt{car()} and \texttt{car\_time()}, and \texttt{dagar()} and \texttt{dagar\_time()}. The \texttt{car()} and \texttt{car\_time()} terms support proper and Leroux CAR specifications \citep{banerjee2014hierarchical,LerouxLeiBreslow1999}. A spatial adjacency graph is supplied to each function and can be constructed using \texttt{car\_graph()}. The \texttt{car\_time()} term fits a separable model in which a time column defines the temporal support; the temporal component can use either an ordered-support AR(1) precision or a continuous-time exponential correlation with a temporal decay parameter. The \texttt{dagar()} term uses the same graph support as \texttt{car()}, but orients the graph according to a user-specified ordering and applies the ordered DAGAR construction \citep{DattaDAGAR2019}. The \texttt{dagar\_time()} term combines this ordered DAGAR spatial precision with the same temporal precision options used by \texttt{car\_time()}.

Residual precision models are specified separately from structured process terms. The \texttt{resid()} term allows Gaussian residual variation to depart from the homoskedastic default through fixed row-specific precisions, group-specific residual variances, and direct-estimate variance models. This is particularly useful for area-level small-area estimation, where the response may be a direct estimate with an associated design-based variance. Formula offsets are supported through \texttt{offset()} and enter as known additive linear-predictor components in fitting, fitted values, recovery, and prediction.

The output of \texttt{predict()} is a matrix or array of posterior draws rather than a single point prediction. Posterior predictive inference proceeds by applying any required transformation or aggregation to each retained draw and then summarizing the resulting sample \citep{kerman2007manipulating}. This convention supports domain-level reporting and summaries of change while propagating uncertainty from model fitting, latent-process recovery, prediction, and aggregation through to the final reported quantities.

\section{Software context}
\label{sec:software-context}

The R ecosystem contains extensive spatial and Bayesian modeling tools, and \texttt{stLMM} complements several approaches. Stan and NIMBLE are general probabilistic programming frameworks in which users can specify spatial and space-time models, including models with custom precision matrices, and write the associated prediction and aggregation code \citep{CarpenterEtAl2017Stan,deValpineEtAl2017NIMBLE}. R-INLA provides approximate Bayesian inference for latent Gaussian models, with mature support for areal effects, stochastic partial differential equation (SPDE) spatial fields, and spatial and space-time extensions \citep{RueMartinoChopin2009INLA,LindgrenRueLindstrom2011SPDE,BakkaEtAl2018INLAReview}.

Other packages provide specialized spatial and ecological workflows. \texttt{sdmTMB} uses \texttt{TMB} and SPDE-based Gaussian Markov random fields for spatial and spatiotemporal generalized linear mixed models, including smooth terms, spatially varying coefficients, and hurdle models \citep{KristensenEtAl2016TMB,AndersonEtAl2025sdmTMB}. \texttt{FRK} uses basis functions for spatial and space-time prediction, with explicit support for change-of-support problems \citep{ZammitMangionCressie2021FRK}. \texttt{spBayes} and \texttt{spNNGP} provide Bayesian point-referenced models, with \texttt{spNNGP} focusing on scalable NNGP implementations \citep{FinleyBanerjeeGelfand2015SpBayes,FinleyDattaBanerjee2020SpNNGP}. \texttt{CARBayes} and \texttt{geostan} provide areal regression models \citep{Brown2015CARBayes,Donegan2022Geostan}. Ecology-focused packages such as \texttt{hSDM}, \texttt{Hmsc}, \texttt{spOccupancy}, and \texttt{spAbundance} offer model classes for species distributions, communities, occupancy, and abundance, often with spatial random effects \citep{Vieilledent2023hSDM,TikhonovEtAl2020Hmsc,DoserEtAl2022SpOccupancy,DoserEtAl2024SpAbundance}.

These tools differ in both spatial representation and inference. For example, SPDE approaches, as used by R-INLA and \texttt{sdmTMB}, approximate continuous fields with sparse Gaussian Markov random fields on meshes and project observations onto those meshes. \texttt{stLMM} represents point-referenced dependence through covariance functions or NNGP neighbor conditionals, and areal dependence through CAR or DAGAR graph precisions. Table~S3 in Appendix~S3 (Supporting Information) compares spatial supports, model terms, response distributions, inference, prediction, and aggregation across selected packages.

\texttt{stLMM}'s focus is a common Bayesian formula, recovery, and prediction workflow across point, areal, temporal, and space-time supports. This is useful in ecological monitoring when users need to move between individual measurements and direct estimates, account for their different residual variances, predict at unsampled supports, and aggregate posterior draws to reporting domains within one framework.

\section{Illustration across spatial supports}
\label{sec:simulation}

A monitoring program may release plot measurements with exact coordinates, provide only reporting-area identifiers, or publish direct estimates for those areas. These three settings separate the support of the observations from that of the spatial process: individual measurements can be associated with location-specific or shared areal effects, while direct estimates are modeled at the area level. We use one simulated population and sample to show how the formula interface accommodates these settings and how predictions address a common reporting question.

\begin{figure}[p]
  \centering
  \includegraphics[width=\textwidth,height=0.76\textheight,keepaspectratio]{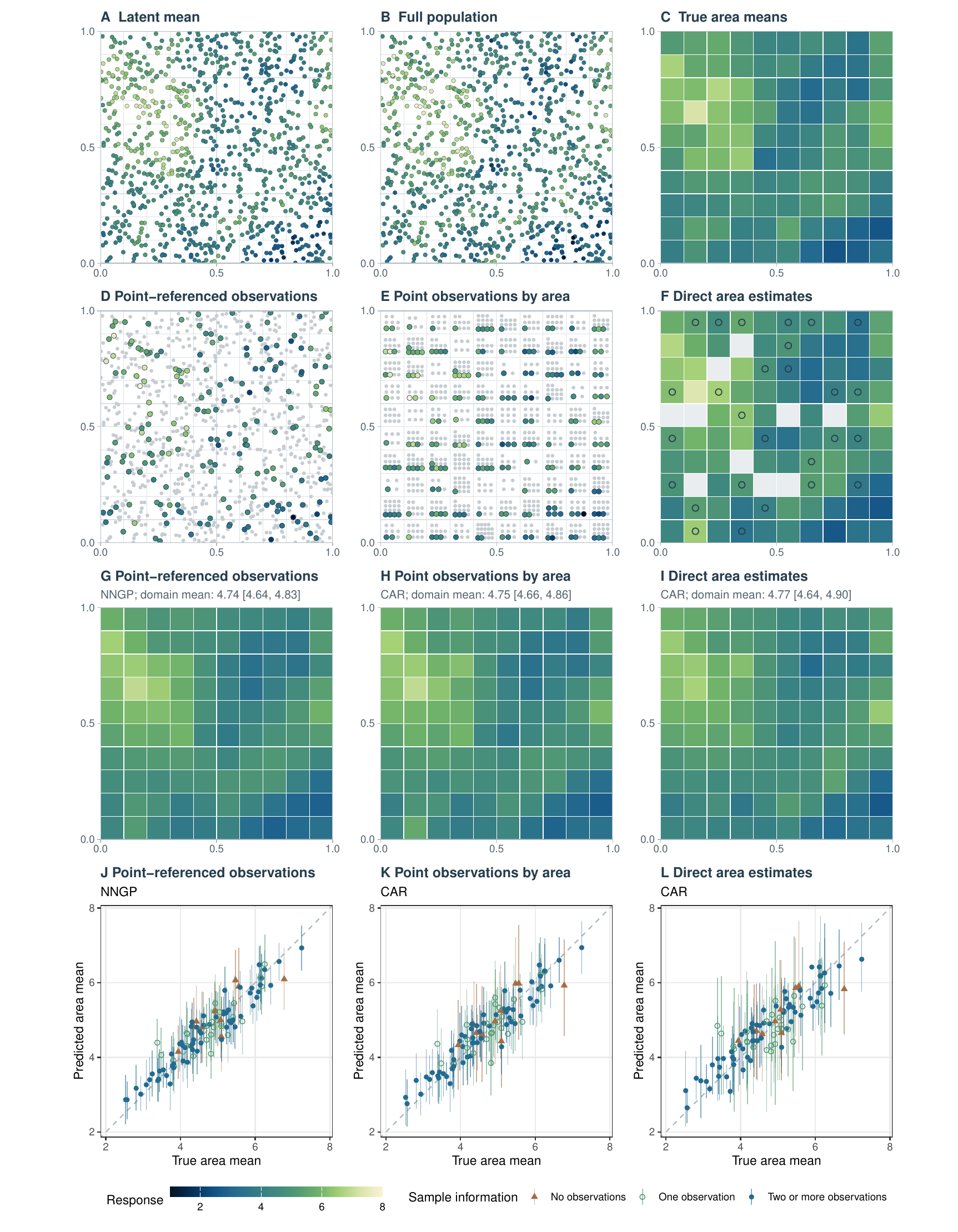}
  \caption{One population, three observation settings, and a common reporting target. A shows latent means at all 1,000 population locations; B shows the full realized population after adding unit-level variation; C averages B within areas. The lower columns follow point-referenced observations (NNGP), point observations by area (CAR), and direct area estimates (CAR). D--F show information from the same 200 sampled units. Gray points denote unsampled units; E arranges observations from the lower left within each area, with positions indicating area membership only. Open circles in F flag singleton estimates lacking sampling variances. G--I map posterior mean area predictions; subtitles give whole-population posterior means and 95\% credible intervals. The true whole-population mean is 4.71. J--L compare predictions with the population area means in C, with 95\% intervals and symbols distinguishing unsampled, singleton, and other sampled areas.}
  \label{fig:simulation-supports}
\end{figure}

The first row of Fig.~\ref{fig:simulation-supports} describes the full population, before sampling. We place 1,000 units randomly in a unit square divided into a $10\times10$ grid of reporting areas. Panel A shows an intercept plus one spatial GP realization at every population location. Adding independent unit-level variation gives the realized responses in B: these are the values for all 1,000 population units. Panel C averages those values within each area. For this synthetic response, the targets are the mean response across all units in each reporting area and across the whole population, including the unit-level variation.

\begin{samepage}
The second row shows the information available from a random sample of 200 units. The same sampled responses underlie all three columns, but their spatial information and level of aggregation differ. Each model includes an intercept and one spatial process term. In D and E, gray points show the 800 unsampled units for which responses will be predicted. The abbreviated calls below highlight the changes in specification; Appendix~S1 (Supporting Information) gives the generating parameters, priors, complete fitting calls, prediction and aggregation details, and diagnostics. The accompanying R script provides the full workflow.
\par
\end{samepage}

\begin{samepage}
\paragraph{Point-referenced observations (D).} Exact coordinates locate each sampled response within the population. An NNGP approximates the generating GP, using distances between locations to represent dependence. The data frame \texttt{point\_data} has 200 rows and three columns: \texttt{coord\_x}, \texttt{coord\_y}, and \texttt{y}.
\nopagebreak[4]

\noindent
\begin{minipage}{\linewidth}
\begin{lstlisting}[style=stlmmR]
fit_nngp <- stLMM(
  y ~ nngp(coord_x, coord_y, m = 15, cov_model = "exp"),
  data = point_data, ...)
\end{lstlisting}
\end{minipage}
\par
\end{samepage}

\begin{samepage}
\paragraph{Point observations by area (E).} Individual responses and area identifiers remain available, but exact coordinates are withheld. Points in E are arranged schematically to show area membership. A CAR effect is shared by all observations in an area, with dependence between areas that share an edge in the neighbor graph. Thus, multiple measurements map to the same latent process node while retaining their separate residuals. The data frame \texttt{point\_in\_area\_data} has 200 rows and three columns: \texttt{unit\_id}, \texttt{area\_id}, and \texttt{y}. The graph \texttt{g} contains all 100 areas, including those with no observations.
\nopagebreak[4]

\noindent
\begin{minipage}{\linewidth}
\begin{lstlisting}[style=stlmmR]
fit_unit_car <- stLMM(
  y ~ car(area_id, graph = g, car_model = "leroux"),
  data = point_in_area_data, ...)
\end{lstlisting}
\end{minipage}
\par
\end{samepage}

\begin{samepage}
\paragraph{Direct area estimates (F).} Only the sample mean $\widehat y_a$ and estimated sampling variance $\widehat v_a=s_a^2/n_a$ from each reporting area $a=1,\ldots,100$ enter the model. Here $n_a$ is the number of sampled units in area $a$, and $s_a^2$ is the sample variance of their responses. The same CAR graph represents dependence among area means, while \texttt{resid()} treats the sampling variances as fixed, as in the classical FH model. This replaces the common residual variance estimated in the two individual-response models. The data frame \texttt{direct\_area\_data} has 100 rows and three columns: \texttt{area\_id}, \texttt{y\_hat} (the direct estimate), and \texttt{v\_hat} (its estimated sampling variance); areas with missing direct estimates have \texttt{y\_hat = NA}.
\nopagebreak[4]

\noindent
\begin{minipage}{\linewidth}
\begin{lstlisting}[style=stlmmR]
fit_direct_car <- stLMM(
  y_hat ~ car(area_id, graph = g, car_model = "leroux") +
    resid(model = "fixed", variance = v_hat),
  data = direct_area_data, ...)
\end{lstlisting}
\end{minipage}
\par
\end{samepage}

\begin{samepage}
The sample leaves 10 areas unsampled and 27 with just one observation. These singleton responses contribute to the first two models, but their area sampling variances cannot be estimated. Open circles in F identify their means, which are omitted from the direct-estimate likelihood, leaving 63 usable responses. The CAR graph retains these areas, and their effects can be recovered through connections to observed areas.
\par
\end{samepage}

The third row returns all three analyses to the common reporting target in C. For the individual-response models, we retain sampled values, predict unsampled responses including unit-level variation, and average each posterior draw within areas (G--H). For the direct-estimate model, the recovered intercept plus CAR effect supplies draws of each area mean (I). Weighting area draws by population counts gives whole-population means and intervals, reported above G--I. All three intervals contain the true whole-population mean of 4.71. The fourth row compares predicted area means with the known population area means: each point in J--L represents one area, vertical bars give 95\% credible intervals, and the diagonal indicates agreement. Symbols distinguish unsampled, singleton, and other sampled areas, showing how prediction and uncertainty extend to areas with little or no local information.

All three analyses recover the broad spatial pattern in this realization. Predictions from point-referenced observations and point observations by area are similar, while the analysis of direct area estimates gives more variable area predictions and wider intervals on average. These differences reflect both the available information and modeling choices, including the NNGP or CAR representation and treatment of sampling variances. One realization does not establish comparative accuracy or interval calibration; it demonstrates a common specification and posterior-draw workflow across observation and process supports. Adding time extends these mappings to location-time or area-time supports; the Washington application below illustrates the latter.

\section{Example application: Washington county biomass}
\label{sec:application}

The \href{https://finleya.github.io/stLMM/articles/index.html}{Washington county biomass article series} on the package's \texttt{pkgdown} site \citep{PkgdownPackage} applies this workflow to Forest Inventory and Analysis (FIA) data from the USDA Forest Service, with tree canopy cover (TCC) as auxiliary information. Here we summarize the \href{https://finleya.github.io/stLMM/articles/wa-direct-car-time-scaled-variance-tcc-svc.html}{direct-estimate CAR-time example}, which combines temporal borrowing, a spatially varying ecological relationship, and a more flexible residual model.

The response rows are county-year direct estimates of live aboveground biomass density. Let $\widehat y_i$ be the direct estimate for county $a_i$ and year $t_i$, with estimated sampling variance $\widehat v_i$, sample size $n_i$, and standardized county mean tree canopy cover $x_{a_i,t_i}$. The article fits
\begin{equation}
  \widehat y_i =
  \beta_0 + \{\beta_1 + u(a_i)\}x_{a_i,t_i} + w(a_i,t_i) + e_i,
  \qquad
  e_i \sim N(0,\tau_i^2).
\end{equation}
The spatially varying coefficient $u(a)$ is a county-level CAR process that lets the TCC-biomass association vary across Washington. The county-year term $w(a,t)$ is a separable CAR-time process that smooths direct estimates over neighboring counties and adjacent years. The direct-estimate variances enter through a scaled residual model,
\begin{equation}
  \log(\tau_i^2) =
  \log(\kappa) +
  \omega_i\log(\widehat v_i) +
  (1-\omega_i)\log(\tau_0^2),
  \qquad
  \omega_i = \frac{n_i}{n_i + c},
\end{equation}
so the supplied sampling variances define relative precision while the model can recalibrate their overall scale and shrink low-information rows toward a common variance.

The formula combines these terms as follows (priors and sampler controls omitted). The county graph \texttt{g} is created from county polygons, with bridge edges for island components. Missing direct estimates remain as county-year prediction supports. The code uses \texttt{direct\_biomass}, \texttt{direct\_biomass\_vhat}, and \texttt{county\_mean\_tcc\_scaled} for $\widehat y_i$, $\widehat v_i$, and $x_{a_i,t_i}$, respectively.

\begin{minipage}{\linewidth}
\begin{lstlisting}[style=stlmmR]
fit <- stLMM(
  direct_biomass ~
    county_mean_tcc_scaled +
    county_mean_tcc_scaled:car(county_fips, graph = g,
                               car_model = "leroux") +
    car_time(county_fips, year, graph = g,
             car_model = "leroux") +
    resid(model = "scaled",
          variance = direct_biomass_vhat,
          n = n,
          shrinkage = 10),
  data = direct_estimates,
  ...
)

rec <- recover(fit, sub_sample = posterior_sub_sample)
theta_draws <- fitted(rec, summary = FALSE)
\end{lstlisting}
\end{minipage}
\par

\begin{figure}[htbp]
  \centering
  \includegraphics[width=\textwidth,trim=0 1.25cm 0 1cm,clip]{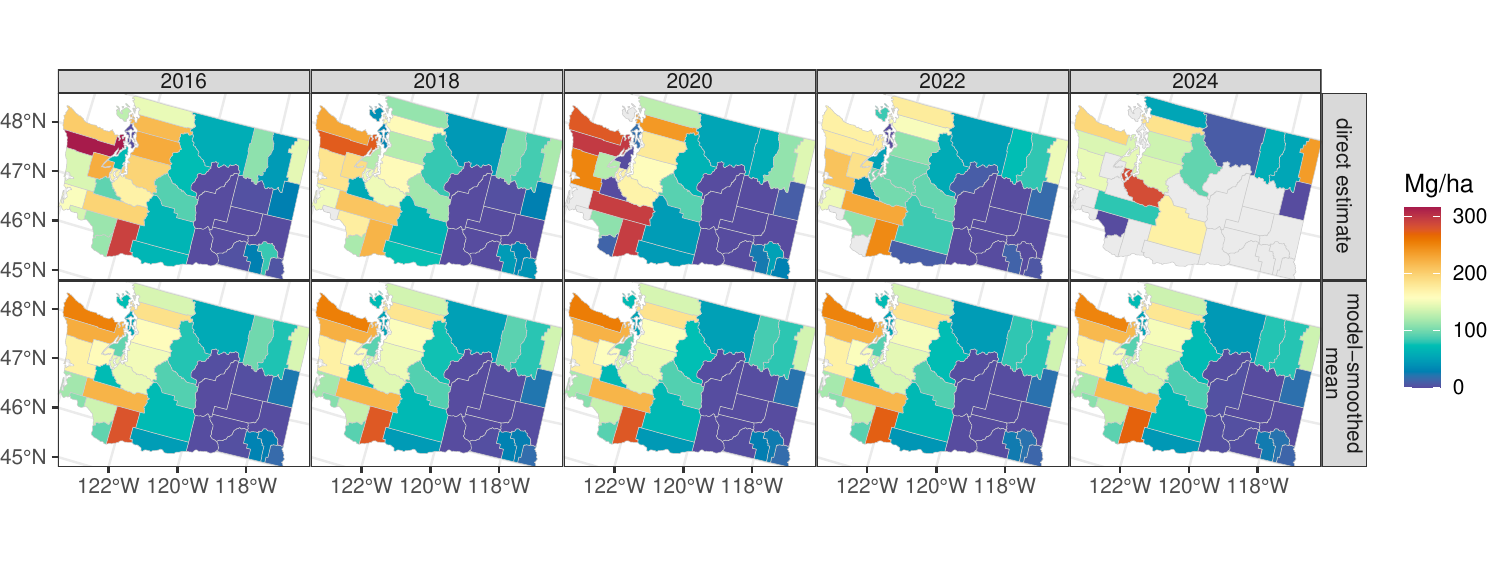}
  \caption{Washington county-year biomass direct estimates and posterior means from the scaled-variance CAR-time model with a spatially varying TCC coefficient. Gray counties have no model-ready direct estimate in that year.}
  \label{fig:wa-direct-smoothed}
\end{figure}

Recovery provides draws of $\theta_{a,t}=\beta_0+\{\beta_1+u(a)\}x_{a,t}+w(a,t)$, the latent county-year biomass means. Fig.~\ref{fig:wa-direct-smoothed} compares their posterior means with the direct estimates. Estimates are available across the retained county-year support, borrowing information through county adjacency, temporal correlation, TCC, and direct-estimate precision. This extends the model for direct area estimates in the simulation while allowing the TCC--biomass relationship to vary among counties.

The full article provides the data, graph construction, priors, sampler settings, diagnostics, and reproducible Quarto source. It also maps the varying TCC coefficient and shows county trajectories with uncertainty. Related articles develop unit-level and other direct-estimate models, providing an extended ecological tutorial alongside the simulation and timing supplements supplied here.

\section{Recommended use and resources}
\label{sec:conclusions}

\texttt{stLMM} provides a unified Bayesian mixed-model interface for ecological applications that require spatial, temporal, or space-time borrowing of strength and posterior prediction beyond the observed sampling locations or reporting units. Its main contribution is not a new model for a single support type, but a common specification and computation layer for related latent process models. This makes it possible to combine iid and structured terms, move between point-referenced and areal supports, and carry posterior uncertainty through fitted values, latent-process recovery, prediction, and user-defined summaries.

The collapsed sampler is also the package's main tradeoff. It provides a common sparse-matrix implementation, but it is not always faster than model-specific latent-variable samplers. Gaussian fits defer latent draws to recovery, whereas augmented non-Gaussian fits retain them within the chain. Sparse Cholesky factorization, NNGP neighbor construction, likelihood augmentation, and posterior prediction can dominate runtime depending on the model and support size. Specialized software may be preferable when one model class is known in advance and speed for that class is the overriding criterion. Flexibility, comparable model specification, and posterior prediction across supports remain the primary aims.

Timing and memory experiments using larger versions of the three illustrative settings in Section~\ref{sec:simulation} provide practical guidance (Appendix~S2, Supporting Information). For point-referenced NNGP models, problems with roughly 100,000 or fewer locations offer a more practical starting range; larger problems are possible, but sampling near a million locations is substantially more demanding. Areal models can accommodate many more observations through shared CAR effects, and our trials extended to approximately five million areas. These short trials measure sampling costs, not time to converged inference. They used an intercept and one spatial process term; costs depend on observation count, process dimension, graph structure, and hardware, and can increase substantially with additional spatial or space-time terms.

The package vignettes introduce individual model terms and diagnostics, while the companion \texttt{pkgdown} site provides a longer-form article series, rendered examples, implementation notes, and the Washington county biomass SAE series summarized in Section~\ref{sec:application}. Additional website articles provide simulation-based parameter-recovery checks for target model classes, validation comparisons with \href{https://finleya.github.io/stLMM/articles/a14-lme4-validation.html}{\texttt{lme4}} \citep{BatesEtAl2015Lme4} and \href{https://finleya.github.io/stLMM/articles/a13-scottish-lip-cancer-nimble-comparison.html}{\texttt{NIMBLE}} \citep{deValpineEtAl2017NIMBLE} for iid and CAR/DAGAR models, \href{https://finleya.github.io/stLMM/articles/a10-metropolis-blocking-diagnostics.html}{sampler-blocking diagnostics}, and \href{https://finleya.github.io/stLMM/articles/a11-runtime-performance.html}{runtime notes}. An in-depth description of model specifications and the shared sparse-precision computations is provided in the package's \href{https://finleya.github.io/stLMM/articles/a12-modeling-software-details.html}{technical reference}.

\section{Software Availability}
\label{sec:data-software}

\texttt{stLMM} is an open-source R package released under the GNU General Public License (GPL, version 3 or later). The source code, issue tracker, and development documentation are hosted on GitHub (\url{https://github.com/finleya/stLMM}). The package website (\url{https://finleya.github.io/stLMM/}) provides function-reference pages, package vignettes, extended application articles, and reproducible article source files. The package can be installed from CRAN or from source using standard R package tools. The package depends on R version 4.4.0 or later, \texttt{Matrix} version 1.7-0 or later \citep{MatrixPackage}, and \texttt{BayesLogit} version 2.4 or later \citep{Polson2013}.

\texttt{stLMM} version 0.0.5 is available on CRAN at \url{https://CRAN.R-project.org/package=stLMM}. The CRAN release provides the permanent, versioned archive of record for the package; CRAN retains earlier versions in its archive, and the development source and rendered documentation are maintained in the GitHub repository and package website described above.

\section{Data Accessibility}
\label{sec:data-accessibility}

The Washington county biomass example uses publicly available data obtained from the USDA Forest Service FIA program (\url{https://www.fia.fs.usda.gov/}), together with tree canopy cover auxiliary data. The prepared county-year data, analysis source, and figure-generating scripts used in Section~\ref{sec:application} are available in the package's source repository and accompanying article series. The simulation and timing scripts and saved timing trials are supplied as ancillary files.

\section*{Acknowledgments}

The author thanks Sudipto Banerjee, Abhirup Datta, and Paul B. May for contributions to the underlying theory and methods for several model classes implemented in \texttt{stLMM}.

Support was provided by the USDA Forest Service Forest Inventory and Analysis Program, the Rocky Mountain Research Station, and the National Council for Air and Stream Improvement (NCASI) Partnership for Small Area Estimation.

\clearpage
\section*{Supporting Information}
\setcounter{section}{0}
\setcounter{subsection}{0}
\setcounter{figure}{0}
\setcounter{table}{0}
\setcounter{equation}{0}
\renewcommand{\theHsection}{S\arabic{section}}
\renewcommand{\theHsubsection}{S\arabic{section}.\arabic{subsection}}
\renewcommand{\theHfigure}{S\arabic{figure}}
\renewcommand{\theHtable}{S\arabic{table}}
\renewcommand{\theHequation}{S\arabic{equation}}

\renewcommand{\thesection}{Appendix S\arabic{section}}
\renewcommand{\thesubsection}{S\arabic{section}.\arabic{subsection}}
\renewcommand{\thetable}{S\arabic{table}}
\renewcommand{\thefigure}{S\arabic{figure}}
\renewcommand{\theequation}{S\arabic{equation}}
\floatplacement{table}{H}

\section{Simulation vignette: one population, three observation settings}
\label{sec:simulation-details}

This appendix develops the three observation settings in Fig.~\ref{fig:simulation-supports}: point-referenced observations, point observations by area, and direct area estimates. Complete R code for population generation, fitting, prediction, and summaries is supplied in the ancillary file \texttt{simulation.R}; the accompanying README gives reproduction instructions.

\subsection{Population, observations, and targets}

We generate a population of size $N=1{,}000$, with independent uniform locations $\bm{s}_i=(s_{i1},s_{i2})\trans$ in the unit square for $i=1,\ldots,N$. A regular $10\times10$ grid defines reporting areas indexed by $a=1,\ldots,100$. Let $\mathcal{P}_a$ index the population units in area $a$, and let $N_a=|\mathcal{P}_a|$ be its population count. A zero-mean Gaussian process (GP) $w(\bm{s})$ generates spatial variation in the response,
\begin{align}
\mu_i &= \beta_0+w(\bm{s}_i), & y_i&=\mu_i+\epsilon_i,\\
\operatorname{Cov}\{w(\bm{s}_i),w(\bm{s}_j)\}
 &= \sigma^2\exp\{-\phi\|\bm{s}_i-\bm{s}_j\|\}, &
\epsilon_i&\stackrel{\mathrm{iid}}{\sim}N(0,\tau^2).
\end{align}
Here $\mu_i$ is the latent mean at location $\bm{s}_i$, and $y_i$ is the realized response. We use intercept $\beta_0=5$, GP variance $\sigma^2=1$, unit-level variance $\tau^2=0.25$, and decay $\phi=-\log(0.05)/0.5=5.991$. Thus spatial correlation reaches 0.05 at distance 0.5.

A simple random sample of size $n=200$ without replacement leaves 10 areas unsampled, 27 with one observation, and 63 with at least two. All three analyses use this sample. Point-referenced observations supply coordinates and responses; point observations by area supply responses and area identifiers, with exact coordinates withheld; direct area estimates supply each area's sample mean $\widehat y_a$ and estimated sampling variance $\widehat v_a=s_a^2/n_a$, where $n_a$ is the number of sampled units in area $a$ and $s_a^2$ is their sample variance.

For direct area estimates, singleton means have no estimable sampling variance and are set to missing, alongside unsampled areas. The CAR graph retains all areas, including those with missing direct estimates. All 200 individual responses, including those in singleton areas, contribute to the models for point-referenced observations and point observations by area.

\begin{samepage}
The targets are the realized area means $\overline y_a=N_a^{-1}\sum_{i\in\mathcal{P}_a}y_i$ and the whole-population mean $\overline y=N^{-1}\sum_{i=1}^N y_i$. Randomly placed units produce unequal area counts. Whole-domain draws therefore weight areas by $N_a/N$.
\par
\end{samepage}

\begin{samepage}
\subsection{Models, priors, and sampling}

Point-referenced observations are modeled with an exponential nearest-neighbor Gaussian process (NNGP) with 15 neighbors and the default coordinate ordering at the observed locations. The two areal fits use Leroux conditional autoregressive (CAR) effects on the default queen-neighbor graph from \texttt{car\_graph()}, connecting areas that share an edge or corner. The CAR model for point observations by area estimates a common residual variance. The model for direct area estimates uses
\begin{equation}
\widehat y_a\mid\theta_a\sim N(\theta_a,\widehat v_a),
\qquad \theta_a=\beta_0+u_a,
\end{equation}
with a CAR prior on the area-effect vector $\bu=(u_1,\ldots,u_{100})\trans$ and supplied sampling variances treated as fixed. This is a spatial direct-estimate model: its fitted $\theta_a$ represents the area quantity directly. Treating estimated sampling variances as known follows the usual Fay--Herriot specification \citep{FayHerriot1979} and is used here to keep the illustration simple. \citet{Shannon_2024_ERL} use direct-estimate sampling variances and sample sizes to inform inverse-gamma priors on the corresponding variance terms in the hierarchical model. This lets those variances adapt to residual variation after accounting for predictors, avoiding a fixed variance constraint that can otherwise force an additional model-error variance to zero. This construction is available through \texttt{resid()} with \texttt{model = "group"} and \texttt{prior = "shannon"}.
\par
\end{samepage}

All intercept priors are $N(0,10^2)$. Estimated process and residual standard deviations have half-$t$ priors with three degrees of freedom and scale one. The decay prior is uniform on $(1,30)$, and CAR association priors are uniform on $(0,0.99)$. The CAR scale is not the GP marginal variance and has no corresponding generating value in this example.

Fits use \texttt{stLMM} 0.0.5, one OpenMP thread per chain, and four chains of 6,000 returned iterations after separate proposal-calibration warmup. We discard the first 2,000 iterations per chain, use all 16,000 remaining draws for parameter summaries and diagnostics, and retain every tenth draw for prediction (1,600 combined draws). In the complete calls below, \texttt{n\_samples = 6000}, \texttt{n\_chains = 4}, and \texttt{g} is the queen-neighbor graph.

\begin{samepage}
For point-referenced observations, \texttt{point\_data} has 200 rows and three columns: \texttt{coord\_x}, \texttt{coord\_y}, and \texttt{y}. All rows have observed responses; unsampled locations enter at prediction.
\nopagebreak[4]

\begin{minipage}{\linewidth}
\begin{lstlisting}[style=stlmmR]
fit_nngp <- stLMM(
  y ~ nngp(coord_x, coord_y, m = 15, cov_model = "exp"),
  data = point_data,
  priors = list(
    beta = normal(0, 10),
    resid = list(tau_sq = half_t(df = 3, scale = 1)),
    nngp_1 = list(sigma_sq = half_t(df = 3, scale = 1), phi = uniform(1, 30))
  ),
  n_samples = n_samples, chains = n_chains, chain_control = list(seed = 1),
  n_omp_threads = 1, verbose = FALSE
)
\end{lstlisting}
\end{minipage}
\par

\par
\end{samepage}
\begin{samepage}
For point observations by area, \texttt{point\_in\_area\_data} has 200 rows and three columns: \texttt{unit\_id}, \texttt{area\_id}, and \texttt{y}. All responses are observed. Repeated area identifiers map units to shared CAR effects, and \texttt{g} retains all 100 areas, including those without observations. The 800 unsampled units enter through \texttt{predict()}.
\nopagebreak[4]

\begin{minipage}{\linewidth}
\begin{lstlisting}[style=stlmmR]
fit_unit_car <- stLMM(
  y ~ car(area_id, graph = g, car_model = "leroux"),
  data = point_in_area_data,
  priors = list(
    beta = normal(0, 10),
    resid = list(tau_sq = half_t(df = 3, scale = 1)),
    car_1 = list(sigma_sq = half_t(df = 3, scale = 1), rho = uniform(0, 0.99))
  ),
  n_samples = n_samples, chains = n_chains, chain_control = list(seed = 1),
  n_omp_threads = 1, verbose = FALSE
)
\end{lstlisting}
\end{minipage}
\par

\par
\end{samepage}
\begin{samepage}
For direct area estimates, \texttt{direct\_area\_data} has 100 rows and three columns: \texttt{area\_id}, \texttt{y\_hat} (the direct estimate), and \texttt{v\_hat} (its estimated sampling variance). Only 63 responses have usable sampling variances; areas with missing direct estimates have \texttt{y\_hat = NA} and remain in the graph.
\nopagebreak[4]

\begin{minipage}{\linewidth}
\begin{lstlisting}[style=stlmmR]
fit_direct_car <- stLMM(
  y_hat ~ car(area_id, graph = g, car_model = "leroux") +
    resid(model = "fixed", variance = v_hat),
  data = direct_area_data,
  priors = list(
    beta = normal(0, 10),
    car_1 = list(sigma_sq = half_t(df = 3, scale = 1), rho = uniform(0, 0.99))
  ),
  n_samples = n_samples, chains = n_chains, chain_control = list(seed = 1),
  n_omp_threads = 1, verbose = FALSE
)
\end{lstlisting}
\end{minipage}
\par

\par
\end{samepage}
\subsection{Prediction and aggregation}

For the models of point-referenced observations and point observations by area, sampled values are retained and unsampled values are predicted, including their nugget variation. For posterior draw $\ell$,
\begin{equation}\label{eq:area-aggregation}
\overline y_a^{(\ell)}=\frac{\sum_{i\in S_a}y_i+\sum_{i\in U_a}y_i^{(\ell)}}{N_a},
\end{equation}
where $S_a$ and $U_a$ index sampled and unsampled population units in area $a$. Joint NNGP prediction retains spatial dependence needed for uncertainty in aggregates, while CAR predictions share recovered area effects.

For the model of direct area estimates, recovered fitted draws of $\theta_a$ represent the area mean. In every setting, domain draws are $\overline y^{(\ell)}=\sum_{a=1}^{100}(N_a/N)\overline y_a^{(\ell)}$. Aggregation is performed within each joint posterior draw, preserving dependence among areas and propagating uncertainty to the domain mean \citep{kerman2007manipulating}. Posterior means and equal-tailed 95\% intervals summarize the resulting draws.

The following code shows recovery, prediction, and draw extraction. In \texttt{population\_data}, unsampled responses are \texttt{NA}. All three output matrices have posterior draws in rows: \texttt{y\_draws\_nngp} and \texttt{y\_draws\_unit\_car} have columns for the 800 unsampled units, while \texttt{area\_draws\_direct\_car} has columns for the 100 areas. In \texttt{simulation.R}, \eqref{eq:area-aggregation} combines the first two matrices with observed responses; the direct-estimate draws already represent area means.

\begin{minipage}{\linewidth}
\begin{lstlisting}[style=stlmmR]
posterior_sub_sample <- list(start = burn + 1L, thin = thin_prediction)
new_points <- population_data |> filter(is.na(y)) |> select(-y)

# Point-referenced observations.
rec_nngp <- recover(fit_nngp, sub_sample = posterior_sub_sample)
pred_nngp <- predict(
  rec_nngp, newdata = new_points, y_samples = TRUE,
  joint = TRUE, joint_method = "vecchia", pred_m = 15,
  return_w_samples = FALSE
)
y_draws_nngp <- as.matrix(as_samples(pred_nngp, sample = "y",
                                   metadata = FALSE))

# Point observations by area.
rec_unit_car <- recover(fit_unit_car, sub_sample = posterior_sub_sample)
pred_unit_car <- predict(
  rec_unit_car, newdata = new_points |> select(unit_id, area_id),
  y_samples = TRUE, return_w_samples = FALSE
)
y_draws_unit_car <- as.matrix(as_samples(pred_unit_car, sample = "y",
                                       metadata = FALSE))

# Direct area estimates.
rec_direct_car <- recover(fit_direct_car, sub_sample = posterior_sub_sample)
area_draws_direct_car <- as.matrix(as_samples(
  fitted(rec_direct_car, summary = FALSE), metadata = FALSE))
\end{lstlisting}
\end{minipage}
\par

\clearpage
\subsection{Results and diagnostics}

Applying the recovery, prediction, and aggregation steps above to the three fitted models gives posterior summaries for the same 100 population area means. Table~\ref{tbl-area-errors} compares posterior mean predictions with the known population values and summarizes the associated 95\% intervals, complementing Fig.~\ref{fig:simulation-supports}. Table~\ref{tbl-parameters} reports parameter estimates and MCMC diagnostics for these fits.

\begin{table}

\caption{\label{tbl-area-errors}Area-mean errors and 95\% interval summaries over 100 reporting areas. Mean width is the average interval width; coverage is the percentage of reporting areas whose intervals contain the true area mean.}

\centering{

\centering
\begin{tabular}[t]{lrrrr}
\toprule
Observation setting & RMSE & Mean error & Mean width & Coverage (\%)\\
\midrule
Point-referenced observations & 0.350 & 0.020 & 1.283 & 90\\
Point observations by area & 0.372 & 0.032 & 1.543 & 94\\
Direct area estimates & 0.445 & 0.052 & 1.780 & 82\\
\bottomrule
\end{tabular}

}

\end{table}%

\begin{figure}[H]
\centering
\includegraphics[width=\textwidth,height=0.62\textheight,keepaspectratio]{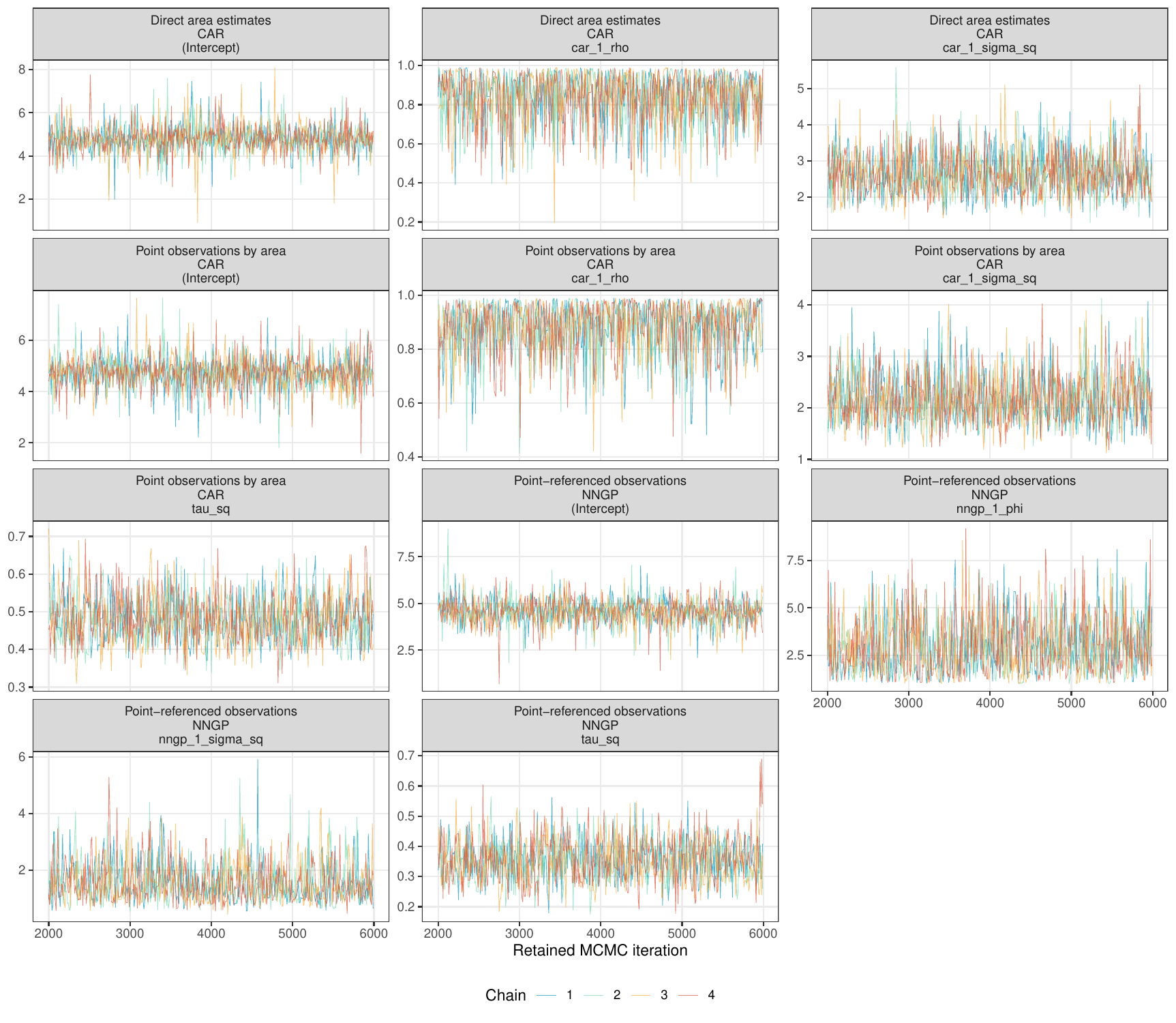}
\caption{Parameter traces after discarding 2,000 iterations from each of four chains. Displayed traces use every tenth draw; numerical diagnostics use all post-burn draws.}
\label{fig:traces}
\end{figure}

\begin{samepage}
The multi-chain method \texttt{summary(fit, burn = 2000)} reports posterior estimates, the Gelman--Rubin potential scale reduction diagnostic ($\widehat R$), and effective sample size (ESS), using \texttt{coda} \citep{PlummerEtAl2006Coda}. The largest parameter $\widehat R$ is 1.007, and the smallest ESS is 1031; Fig.~\ref{fig:traces} shows the chains.
\par
\end{samepage}

\begin{table}

\caption{\label{tbl-parameters}Posterior parameter summaries and coda diagnostics. Generating values are given in the population model; CAR scales describe a different spatial representation.}

\centering{

\centering\small
 
\textbf{Point-referenced observations}\par\smallskip

\begin{tabular}{lrlrr}
\toprule
Parameter & Mean & 95\% interval & R-hat & ESS\\
\midrule
Intercept & 4.576 & (3.118, 5.857) & 1.001 & 15721\\
Residual variance & 0.356 & (0.238, 0.499) & 1.007 & 1284\\
Process scale & 1.565 & (0.705, 3.282) & 1.000 & 1031\\
Decay & 2.975 & (1.104, 6.332) & 1.003 & 1092\\
\bottomrule
\end{tabular}
\par\medskip
\textbf{Point observations by area}\par\smallskip

\begin{tabular}{lrlrr}
\toprule
Parameter & Mean & 95\% interval & R-hat & ESS\\
\midrule
Intercept & 4.745 & (3.411, 6.021) & 1.000 & 18991\\
Residual variance & 0.480 & (0.373, 0.617) & 1.002 & 1270\\
Process scale & 2.182 & (1.397, 3.231) & 1.000 & 1379\\
CAR association & 0.880 & (0.636, 0.987) & 1.000 & 1285\\
\bottomrule
\end{tabular}
\par\medskip
\textbf{Direct area estimates}\par\smallskip

\begin{tabular}{lrlrr}
\toprule
Parameter & Mean & 95\% interval & R-hat & ESS\\
\midrule
Intercept & 4.778 & (3.455, 6.091) & 1.001 & 14887\\
Process scale & 2.659 & (1.707, 3.993) & 1.001 & 1755\\
CAR association & 0.842 & (0.518, 0.985) & 1.001 & 1523\\
\bottomrule
\end{tabular}
\par\medskip

}

\end{table}%

\clearpage
\section{Computational benchmarks}
\label{sec:timing-details}

The ancillary script \texttt{timing.R} reads \texttt{timing\_trials.csv} to reproduce Figs.~\ref{fig:sampling-resources}--\ref{fig:fixed-areas} and their summaries. The \texttt{benchmark/} directory supplies a script and instructions for new trials.

\subsection{Methods}

We measure sampling time and memory across the three observation settings in Appendix S1, fitting an intercept and one spatial process term in each case. The point-referenced model uses an exponential NNGP with 15 neighbors; both areal models use Leroux CAR effects on a regular grid. Responses at random locations in the unit square combine a smooth spatial surface with Gaussian noise. These synthetic datasets provide workloads for assessing computational costs across problem sizes.

Here $n$ counts responses contributing to the likelihood and $q$ counts spatial process effects. For point-referenced observations, $n=q$. Point observations by area can share effects, while direct area estimates contribute at most one response per area. CAR graphs retain areas without usable responses.

Sampling times are scaled from 20 post-warmup iterations to minutes per 1,000 iterations of one chain. They include numerical refactorization but exclude preparation, initialization, warmup, recovery, and prediction. This isolates the recurring computational cost of sampler updates. Peak memory includes data preparation and model fitting.

Trials used \texttt{stLMM} 0.0.5 on an AMD Threadripper 7960X with 24 physical cores and 125.3 GiB RAM, with automatic CHOLMOD fill-reducing ordering. The figures show results with one OpenMP and one BLAS thread. The accompanying code README records the full software environment and trial protocol.

\subsection{Results and interpretation}

\begin{figure}[htbp]
  \centering
  \includegraphics[width=\textwidth]{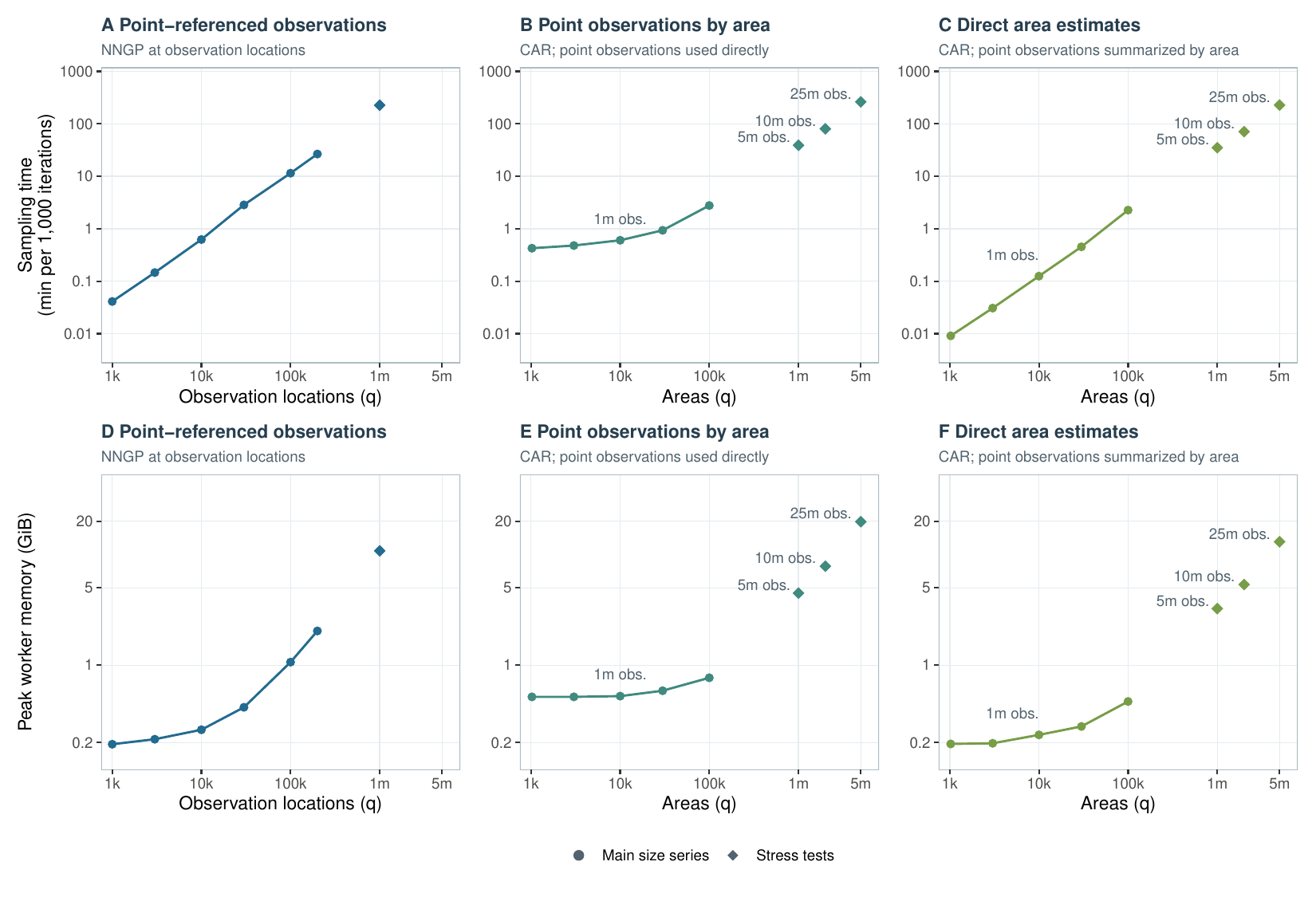}
  \caption{Sampling time (A--C) and peak memory (D--F) as spatial process dimension $q$ increases. Connected circles show the main size series; diamonds show stress tests. Labels in the CAR panels give point-observation counts in millions (m), before aggregation for the direct-estimate models. All axes are logarithmic.}
  \label{fig:sampling-resources}
\end{figure}

Each column of Fig.~\ref{fig:sampling-resources} pairs sampling time above with memory below. Connected circles show the main size series; diamonds show separate stress tests at much larger sizes.

For point-referenced observations (panels A and D), sampling costs rise from 0.6 min per 1,000 iterations at 10,000 locations to 11.5 min at 100,000 locations, where peak memory is 1.06 GiB. These timings suggest roughly 100,000 or fewer locations as a practical starting range for this specification. Costs increase to 26.6 min at 200,000 locations. The million-location diamond reaches 226.8 min per 1,000 iterations. This makes fitting long chains at that size impractical for routine analysis on this machine, even though the measured memory requirement (10.78 GiB) fits comfortably within its RAM.

For point observations by area (panels B and E), every connected-circle case uses one million point observations, while the number of CAR area effects increases from approximately 1,000 to 100,000. For direct area estimates (panels C and F), those same point observations are first summarized by area, so the likelihood contains one response per usable area estimate. At approximately 100,000 areas, sampling costs are 2.8 min per 1,000 iterations and peak memory is 0.77 GiB for point observations by area; direct area estimates require 2.3 min and 0.47 GiB. Both settings therefore accommodate a million underlying point observations across up to about 100,000 areas with modest sampling costs.

\clearpage
The CAR stress tests increase both the number of areas and the underlying point-observation count: five, ten, and 25 million point observations across approximately one, two, and five million areas, respectively. This keeps about five point observations per area on average as the graph grows. The observations enter individually in B and E and as direct area estimates in C and F. At one million areas, costs reach approximately 35--39 min per 1,000 iterations across the two settings. At approximately five million areas, they reach 3.8--4.4 hours, with 13.0--19.8 GiB peak memory. These largest cases demonstrate that the computations can be completed, but their sampling times make routine fitting burdensome.

\begin{figure}[htbp]
\centering
\includegraphics[width=\textwidth]{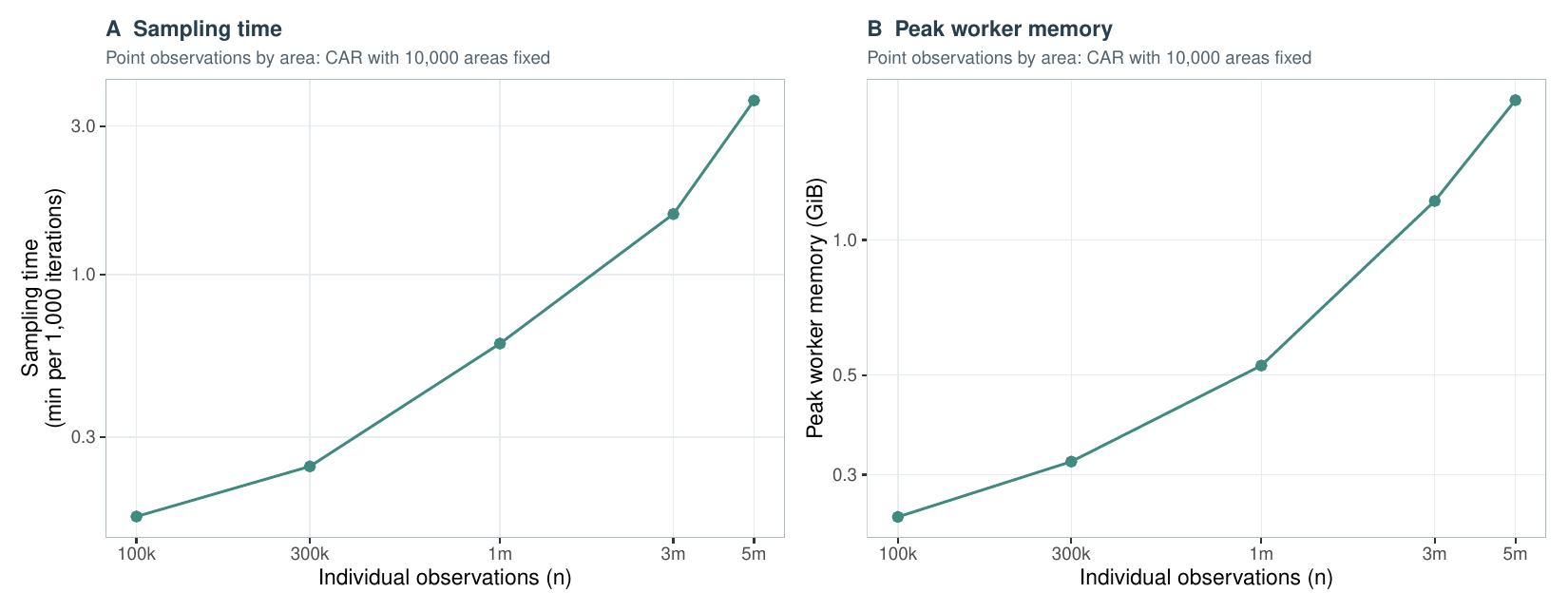}
\caption{Sampling time (A) and peak memory (B) for point observations by area as response count $n$ increases, with $q=10,000$ CAR effects fixed. All axes are logarithmic.}
\label{fig:fixed-areas}
\end{figure}

\clearpage
Figure~\ref{fig:fixed-areas} isolates the cost of adding observations while keeping the spatial process dimension fixed. With 10,000 CAR area effects, increasing point observations by area from 100,000 to five million raises sampling cost from 0.17 to 3.63 min per 1,000 iterations and peak memory from 0.24 to 2.05 GiB. Millions of individual observations can therefore remain practical when they share a moderate number of area effects. The number of areas, as well as the number of observations, determines computational demand.

\clearpage
\section{Comparison of spatial modeling software}
\label{sec:software-comparison}

Table~\ref{tab:software} compares selected spatial modeling workflows. SPDE denotes stochastic partial differential equation; CAR and SAR denote conditional and simultaneous autoregressive models, respectively.

\begin{table}[H]
\centering
\small
\setlength{\tabcolsep}{4pt}
\renewcommand{\arraystretch}{1.15}
\caption{Selected spatial modeling workflows through their standard interfaces. Capabilities depend on the model and response family; entries do not imply that every combination is implemented or rank computational performance. Aggregation describes propagation of uncertainty to reporting domains.}
\label{tab:software}
\begin{tabularx}{\textwidth}{@{}>{\raggedright\arraybackslash}p{0.15\textwidth}>{\raggedright\arraybackslash}p{0.25\textwidth}>{\raggedright\arraybackslash}p{0.24\textwidth}>{\raggedright\arraybackslash}X@{}}
\toprule
Package & Supports, time, and varying coefficients & Responses and inference & Prediction and aggregation\\
\midrule
\texttt{stLMM} & Point and areal; space-time; structured varying coefficients across supports & Gaussian, binomial logit, binary probit, negative binomial; MCMC with collapsed parameter updates & Missing-response supports and new locations; posterior draws for user-defined aggregation\\
\addlinespace
R-INLA & Point via SPDE mesh and areal; space-time; structured varying coefficients & Many families, including Gaussian, binomial, Poisson and negative binomial; approximate Bayesian inference & Missing-response prediction; linear combinations and approximate joint posterior draws\\
\addlinespace
\texttt{sdmTMB} & Point via SPDE mesh; areal CAR/SAR; space-time; structured varying coefficients & Many response families and delta models; Laplace-based likelihood inference, optional \texttt{tmbstan} MCMC & New-data prediction and simulations; area-weighted abundance indices and uncertainty\\
\addlinespace
\texttt{FRK} & Point and areal via basic areal units; space-time; fixed regression effects in standard interface & Gaussian and several non-Gaussian families; expectation-maximization or Laplace-based empirical inference & Predictions on unobserved supports; explicit block/change-of-support prediction and conditional simulations\\
\addlinespace
\texttt{spBayes} & Point; dynamic space-time models; spatially varying coefficients & Gaussian, binomial or Poisson, depending on model; MCMC & New-location posterior prediction; posterior samples for user-defined summaries\\
\addlinespace
\texttt{spNNGP} & Point; spatial NNGP; fixed regression effects in standard interface & Gaussian and binomial; MCMC or conjugate Gaussian models & New-location prediction; posterior samples for user-defined summaries\\
\addlinespace
\texttt{CARBayes}, \texttt{CARBayesST} & Areal; space-time in \texttt{CARBayesST}, including spatially varying time slopes & Gaussian, binomial, Poisson and selected extensions; MCMC & Missing responses in many specifications; posterior samples for user-defined summaries\\
\bottomrule
\end{tabularx}
\par\smallskip
\raggedright\footnotesize
Sources: R-INLA \citep{BakkaEtAl2018INLAReview}; \texttt{sdmTMB} \citep{AndersonEtAl2025sdmTMB,sdmTMBDocumentation2026}; FRK \citep{ZammitMangionCressie2021FRK,SainsburyDaleEtAl2024FRK}; \texttt{spBayes} \citep{FinleyBanerjeeGelfand2015SpBayes}; \texttt{spNNGP} \citep{FinleyDattaBanerjee2020SpNNGP}; CAR packages \citep{Brown2015CARBayes,LeeEtAl2018CARBayesST}. Documentation checked November 2026.
\end{table}

\clearpage
\begingroup\small
\bibliographystyle{abbrvnat}
\bibliography{stLMM_R1}
\endgroup

\end{document}